\documentclass[twocolumn,showpacs,amsmath,amssymb,prx,superscriptaddress,longbibliography]{revtex4-2}

\usepackage[T1]{fontenc}
\usepackage{lmodern}
\usepackage{amssymb}
\usepackage{amsmath}
\usepackage{textgreek}
\usepackage{microtype}
\usepackage {longtable}
\usepackage{mathtools}
\usepackage{array}
\usepackage{bm}
\usepackage{float}
\usepackage{graphicx}
\usepackage{xcolor}
\usepackage{csquotes}
\usepackage{nccmath}
\usepackage[hyperindex,breaklinks]{hyperref}
\newcommand{\ket}[1]{|#1\rangle}

\newcommand \be{\begin{equation}}
\newcommand \ee{\end{equation}}
\newcommand \bea{\begin{eqnarray}}
\newcommand \eea{\end{eqnarray}}
\newcommand \bse{\begin{subequations}}
\newcommand \ese{\end{subequations}}

\begin{document}

\relpenalty=1
\title{Long-range $CC\Phi$ gates via radio-frequency-induced F\"{o}rster resonances}

\author{I.~N.~Ashkarin}
\email{ivan.ashkarin@universite-paris-saclay.fr}
\affiliation {Universit\'e Paris-Saclay, CNRS, Laboratoire Aim\'e Cotton, 91405 Orsay, France}

\author{S.~Lepoutre}
\affiliation {Universit\'e Paris-Saclay, CNRS, Laboratoire Aim\'e Cotton, 91405 Orsay, France}

\author{P.~Pillet}
\affiliation {Universit\'e Paris-Saclay, CNRS, Laboratoire Aim\'e Cotton, 91405 Orsay, France}

%\author{D.~Comparat}
%\affiliation {Universit\'e Paris-Saclay, CNRS, Laboratoire Aim\'e Cotton, 91405 Orsay, France}
%\affiliation {Novosibirsk State  University, Faculty of Physics, 630090 Novosibirsk, Russia}

\author{I.~I.~Beterov}
\affiliation {Rzhanov Institute of Semiconductor Physics SB RAS, 630090 Novosibirsk, Russia}
%\affiliation {Novosibirsk State  University, Faculty of Physics, 630090 Novosibirsk, Russia} 

\author{I.~I.~Ryabtsev}
\affiliation {Rzhanov Institute of Semiconductor Physics SB RAS, 630090 Novosibirsk, Russia}
%\affiliation {Novosibirsk State  University, Faculty of Physics, 630090 Novosibirsk, Russia}

\author{P.~Cheinet}
\affiliation {Universit\'e Paris-Saclay, CNRS, Laboratoire Aim\'e Cotton, 91405 Orsay, France}

\begin{abstract}

Registers of trapped neutral atoms, excited to Rydberg states to induce strong long-distance interactions, are extensively studied for direct applications in quantum computing. Here, we present a novel $CC\Phi$ quantum phase gate protocol based on radio-frequency-induced F\"{o}rster resonant interactions in the array of highly excited $^{87}$Rb atoms. The extreme controllability of interactions provided by RF field application enables high-fidelity and robust gate performance for a wide range of parameters of the atomic system, as well as it significantly facilitates the experimental implementation of the gate protocol. Taking into account finite Rydberg states lifetimes, we achieve an average theoretical gate fidelity of $99.27 \%$ under room-temperature conditions (improved up to $99.65 \%$ in a cryogenic environment), thus showing the protocol compatibility with modern quantum error correction techniques.

\end{abstract}
\maketitle

\section{Introduction}
\label{Sec1}

Reconfigurable arrays of neutral atoms arranged in optical tweezers present a prospective platform for quantum computing \cite{Madjarov2020, Morgado2021, Graham2022, Cuadra2022, Bluvstein2022, Graham2022}. Broad possibilities of interatomic interactions control via Rydberg excitation \cite{Saffman2010, Adams2019, Levine2019, Shi2022, Evered2023}, complemented by long atomic lifetimes in individual traps \cite{Browaeys2020, Schymik2021} and high scalability of quantum registers \cite{Ebadi2022, Scholl2021, Schymik2022, Pause2024, Norcia2024, Manetsch2024} underscore the pivotal virtues of this approach. Recent advances in neutral-atom-based quantum computing include simulations of quantum phase transitions \cite{De2019, Scholl2021, Ebadi2021}, as well as implementation of high-fidelity parallel quantum gates in large-scale registers and generation of entangled multi-qubit states \cite{Graham2019, Levine2019, Evered2023, Shaw2023}.

An outstanding challenge for neutral-atom-based quantum information processing is the implementation of high-fidelity multi-qubit controlled quantum gates. Leveraging multi-qubit operations allows to significantly reduce the total gate count for complex quantum algorithms \cite{Dlaska2022, Zhang2023, Tang2022}, thus paving a way to the practical applications of near-term noisy intermediate-scale quantum (NISQ) devices \cite{Daley2022, Lanthaler2023}. In this regard, a variety of many-body quantum gate schemes has been recently designed, including protocols based on the dipole blockade effect \cite{Isenhower2011, Levine2019, Young2021, Yu2022, Li2022, Jandura2023, Evered2023, Ma2023}, electromagnetically induced transparency \cite{McDonnell2022, Farouk2022} and Rydberg antiblockade \cite{Yin:20, Wu2021, Yang2021, Wu2022}. Nevertheless, utilization of the described approaches is limited by the van der Waals character of the interatomic interaction, which imposes significant constraints on the distances between qubits involved in the gate protocol ($ \sim 2-5$ \textmu m for typical experimental realizations) \cite{Levine2019, Evered2023}. In turn, the ability to perform gates between remote qubits is important for creating entangled states in large quantum registers and realizing inter-register transport of quantum information \cite{Adams2019, Morgado2021}. To circumvent this bottleneck, an alternative technique based on three-body Stark-induced Förster resonant interactions \cite{Faoro2015, Tretyakov2017} has been proposed for the implementation of fast high-fidelity three-qubit quantum gates \cite{Beterov2018a, Ashkarin2022}.
The enhancement of dipole coupling achieved via activation of Förster transitions allows the realization of controlled interactions between distant qubits ($\sim~10-20$ \textmu m), thus providing vast possibilities to achieve increased interconnectivity in the atom-based devices~\cite{Hollerith2022, Saffman2010, Ramette2022}.

%The Förster-based gates implementation complexity due to the technical limits on the control accuracy of the external inducing fields is an essential difficulty for the proposed approach \cite{Ryabtsev2010, Beterov2015, Beterov2016a}. In particular, previously proposed protocols required the application of Stark-switching techniques during the gate process, assuming the external dc electric field control as precise as $\sim 10^{-6}$ V/cm \cite{Beterov2018a}. Significant obstacle is also presented by the strong disturbance of gate fidelity due to interatomic distance variations \cite{Ashkarin2022,Farouk2022}. In this regard, the search for new interaction control and resonance activation methods in order to increase both stability and fidelity of multi-qubit Förster-based gates poses an important challenge.

The Förster-based gates sensitivity to accuracy level of the external inducing fields is an essential difficulty for the proposed approach \cite{Ryabtsev2010, Beterov2015, Beterov2016a}. In particular, previously designed protocols required changing the resonance-inducing DC electric field from a non-resonant value to the exact resonance during the gate process by applying Stark-switching techniques, while assuming the external field control as precise as $\sim 10^{-6}$ V/cm \cite{Beterov2018a}. Significant obstacle is also presented by the strong disturbance of gate fidelity due to interatomic distance variations \cite{Ashkarin2022,Farouk2022}. In this regard, the search for new interaction control and resonance activation methods in order to increase both robustness and fidelity of multi-qubit Förster-based gates poses an important challenge.

One of the effective techniques of Förster resonance induction is the application of external radio-frequency (RF) radiation. Previously, RF-induced two-body resonances were thoroughly investigated in Rydberg ensembles \cite{Tauschinsky2008, Ditzhuijzen2009, Tretyakov2014, Yakshina2016, Lee2017}. High-fidelity two-qubit gate protocols, robust to the experimental imperfections were also proposed based on this approach \cite{Huang2018, Beterov2018}. Thus, extended research on RF induction of many-body resonant interactions presents an interest for potential applications in quantum computing.

In this paper, we propose a protocol for arbitrary doubly-controlled $CC\Phi$ quantum phase gate based on a novel RF-induced three-body F\"{o}rster resonant interaction. The developed protocol exhibits high fidelity for arbitrary phase values ($\sim 99.2 - 99.7 \%$). We explore three-body resonant transfers in ordered arrays of three $^{87}$Rb atoms and show that under the action of a composite electric field consisting of a static (DC) and a dynamical (AC) parts, these transfers acquire replicas that give access to coherent population and phase dynamics.  We demonstrate that RF pulse duration along with the radiation frequency and amplitude provide sufficient interaction control parameters to supervise the gate behaviour. Another important feature of the proposed protocol is the possibility to activate the gate solely by applying an RF pulse, thus eliminating the necessity to use Stark-switching techniques and significantly increasing the potential robustness of the gate to electric field deviations compared to the previous proposals \cite{Beterov2018a, Ashkarin2022}.

\section{Resonance model}
\label{Sec2}
Previously, we have investigated many-body Förster resonant interactions in systems of ultracold Rydberg atoms both theoretically \cite{Ryabtsev2018, Beterov2018, Cheinet2020} and experimentally \cite{Gurian2012, Faoro2015, Tretyakov2017}. It was shown that three-body resonances can be represented as two concurrent two-body dipole transitions \cite{Ryabtsev2018, Beterov2018a} between collective three-atom states of the form $\ket{n_{1} l_{1} j_{1} \left(m_{j1}\right) ;n_{2} l_{2} j_{2} \left(m_{j2}\right) ;n_{3} l_{3} j_{3} \left(m_{j3}\right)} $. In turn, the dipole-dipole interaction (DDI) operator between two neighboring atoms positioned along the quantization axis (Z) can be expressed as~\cite{Walker2008}:
\begin{equation}
\label{eq1}
    \begin{aligned}
V_{dd} &= \frac{e^{2} }{4\pi \varepsilon _{0} R^{3} } \left(\textbf{a}\cdot \textbf{b}-3a_{z} b_{z} \right)=\\
&= -\frac{\sqrt{6} e^{2} }{4\pi \varepsilon _{0} R^{3} } \sum _{q=-1}^{1}C_{1q \; 1-q}^{20} a_{q} b_{-q}.  
\end{aligned}
\end{equation}
Here $\varepsilon_0$ is the vacuum dielectric constant; $e$ is the electron charge; \textbf{a} and \textbf{b} are the vectorial positions of the Rydberg electrons and $C_{1q \; 1-q}^{20}$ represent the Clebsch–Gordan coefficients. The radial matrix elements of the dipole moment are calculated using a quasiclassical approximation~\cite{Kaulakys1995}. Note that due to positioning of atoms along the Z axis, the operator (Eq. \ref{eq1}) only couples collective atomic states with $\Delta M = 0$, where $M= \sum_i m_i$ is the total momentum projection of the state \cite{Ryabtsev2018, Beterov2018a}.
\begin{equation}
    \begin{aligned}
\label{eq2}
\ket{nP_{3/2}}^{\otimes 3} \longleftrightarrow &\ket{nS_{1/2};(n+1)S_{1/2};nP_{3/2}} \longleftrightarrow \\ &\ket{nS_{1/2};nP_{1/2};(n+1)S_{1/2}}
\end{aligned}
\end{equation}

We consider three-body fine-structure-state-changing (FSSC) F\"{o}rster resonant transitions of the form (Eq.~\ref{eq2}) in a linear ensemble of three $^{87}$Rb atoms, isolated in individual optical tweezers at a distance $R$ from each other along the direction of the control electric field~\cite{Cheinet2020}. Here, AC Stark shifts induced by trapping lasers are neglected and the atoms are considered to be fixed spatially during the whole interaction time (unless specified otherwise). Being initially excited into identical Rydberg states $\ket{nP_{3/2}}$, the atoms pass into a collective state $\ket{nS_{1/2};nP_{1/2};(n+1)S_{1/2}}$ as a result of a Stark-induced resonant transfer. A distinctive feature of this scheme is that one of the atoms passes to the $nP$ state with $j=1/2$ during non-resonant two-body $SP$ excitation hopping. The negative zero-field energy defect of the two-body transfer $\ket{nP_{3/2}}^{\otimes 2} \leftrightarrow \ket{nS;(n+1)S}$, in turn, guarantees the absence of two-body resonant interactions in the vicinity of the considered three-body transition. Note that the described resonances (Eq. \ref{eq2}) exist for arbitrary value of the principal quantum number $n$, which thus can be chosen depending on the experimental requirements.

%Using RF induction, resonances (Eq. \ref{eq2}) can be activated for arbitrary values of DC electric field. We define the F\"orster energy defect $\Delta_F$ as the difference between the energies of the final and initial collective states. The resonance induction results from the compensation of the Förster defect by the energy provided by the RF photons of frequency $\omega$ under the condition $\Delta_F=m \omega$ for integer $m$.

We define the F\"orster energy defect $\Delta_F$ as the difference between the energies of the final and initial collective states in external DC electric field. Using additional RF pulse, resonances (Eq. \ref{eq2}) can be activated for arbitrary values of DC field. The resonance induction results from the compensation of the Förster defect by the energy provided by the RF photons of frequency $\nu$ under the condition $\Delta_F=s \nu$ for a number of photons $s$.

Numerical approach to the simulation of resonant interactions induced by a time-dependent periodic electric field was analysed in detail in \cite{Ditzhuijzen2009}. Following the course of this work in combination with our previous study~\cite{Beterov2018a}, we solve numerically the non-Hermitian Hamiltonian based Schr\"{o}dinger equations for the complex amplitudes of the collective basis states taking into account Rydberg lifetimes~\cite{Beterov2009, Beterov2018a}. The basis is represented by combinations of magnetic sublevels of the corresponding $nS$, $(n+1)S$ and $nP$ states. For simplicity, we consider an open system and neglect the population redistribution due to decay processes, assuming that finite lifetimes only lead to irrecoverable population losses.

During the simulation, we take into account the dipole-dipole interatomic interactions \cite{Walker2008}, as well as the interaction of atoms with an external electric field $F = F_S+F_{RF}\cos\left(2\pi \nu t\right)$ \cite{Ditzhuijzen2009}. Here $F_S$ denotes the static part of electric field (DC field), while $F_{RF}$ is the amplitude of RF field (AC field), polarized linearly along the DC field component \cite{Sevincli2014}. We consider a range of small electric fields, which do not lead to significant mixture between low-orbit states, thus allowing us to deal with bare zero-field states. The simple form of quadratic Stark shifts $\sim~\alpha_{nL}^m F^2/2$ is assumed for all Rb Rydberg states included in our model.

To provide a clear analytical description of our results, we present additional simulations based on Floquet approach \cite{Ditzhuijzen2009, Sevincli2014, Yakshina2016}. In a composite "DC+AC" electric field, the Rydberg states wavefunctions $\Psi_{nL}^m(\textbf{r},t)$ are described as compositions of an infinite number of Floquet sidebands with relative amplitudes $a_{nL,s}$. This leads to the formation of an infinite number of lines in the spectrum of individual Rydberg atoms, separated by the frequency of the applied field $\nu$.
%\begin{equation}
 %   \begin{aligned}
%\label{eq3}
 %   \Psi_{nL}(\textbf{r},t) &= \psi_{nL}(\textbf{r}) e^{i\alpha_{nL}(F^2_S + F^2_{RF}/2)t/2} \cdot \\ &\cdot \sum_{m=-\infty}^{\infty}a_{nL,m}e^{im\nu t}
%\end{aligned}
%\end{equation}

\begin{equation}
    \label{eq3}
    \Psi_{nL}^m(\textbf{r},t) = \psi_{nL}^m(\textbf{r}) e^{i\alpha_{nL}^m(F^2_S + F^2_{RF}/2)t/2} \sum_{s=-\infty}^{\infty}a_{nL,s}e^{is\nu t}
\end{equation}

\begin{equation}
    \label{eq4}
    a_{nL,s} = \sum_{k=-\infty}^{\infty}J_{s-2k}\left(\frac{\alpha_{nL}^mF_{S}F_{RF}}{\nu}\right) J_{k}\left(\frac{\alpha _{nL}^mF_{RF}^2}{8\nu}\right)
\end{equation}

%\begin{equation}
 %   \begin{aligned}
%\label{eq4}
 %   a_{nL,m} = \sum_{k=-\infty}^{\infty}&J_{m-2k}\left(\frac{\alpha_{nL}F_{S}F_{RF}}{\omega}\right) \cdot \\ &\cdot J_{k}\left(\frac{\alpha _{nL}F_{RF}^2}{8\omega}\right)
%\end{aligned}
%\end{equation}
According to \cite{Yakshina2016}, the intersection of collective states Floquet sidebands in the external electric field results in the appearance of a Förster resonant transition. Note that both DC and AC components are necessary to effectively drive the resonances according to the Floquet model, since at $F_S=0$ the $a_{nL,s}$ coefficients are null for odd $s$ sidebands in Eq. (\ref{eq4}), while the even sidebands remain weak \cite{Ditzhuijzen2009}.

\section{Three-body resonance simulation}

This section provides a detailed description of our simulation results for three-body F\"{o}rster resonant interactions (Eq. \ref{eq2}) induced by RF radiation in a three-atom system. According to the previous studies \cite{Beterov2018a, Cheinet2020, Pham2020}, a good compromise between robustness to small fluctuations of DC electric field and individual Rydberg lifetimes during the resonance process can be demonstrated for $65\leq n \leq 85$ in $^{87}$Rb ensemble. Following the course of our previous work \cite{Ashkarin2022}, we choose $n=70$ to expose the described resonances. To reduce the number of basis states, we limit the collective state spectra within $\pm~2$~GHz range relative to the $\ket{70P_{3/2}(m_j=1/2)}^{\otimes 3}$ initial state energy, thus considering only atomic states with a momentum value $l~\le~1$. Also, due to the selection rule $\Delta M = 0$, raised by the linear configuration of the atomic register, we can exclude states whose total momentum projection differs from $M$ of the initial state \cite{Ryabtsev2018}. Thus, we significantly reduce the basis of states under consideration, down to $\sim 60$ collective three-atom states, presented by combinations of the magnetic sublevels \mbox{$\ket{70S_{1/2} (m_j=\pm1/2)}$}, \mbox{$\ket{71S_{1/2}(m_j=\pm1/2)}$}, \mbox{$\ket{70P_{1/2}(m_j=\pm1/2)}$}, $\ket{70P_{3/2}(m_j=\pm1/2;\pm3/2)}$.
%%%%%%%%%%%%%%%%%
\begin{eqnarray}
\label{eq5}
&\ket{70P_{3/2}}^{\otimes 3} \to \ket{70S_{1/2};70P_{1/2};71S_{1/2}}
\end{eqnarray}
%%%%%%%%%%%%%%%%

\begin{center}
\begin{figure}[!ht]
\center
\includegraphics[width=\columnwidth]{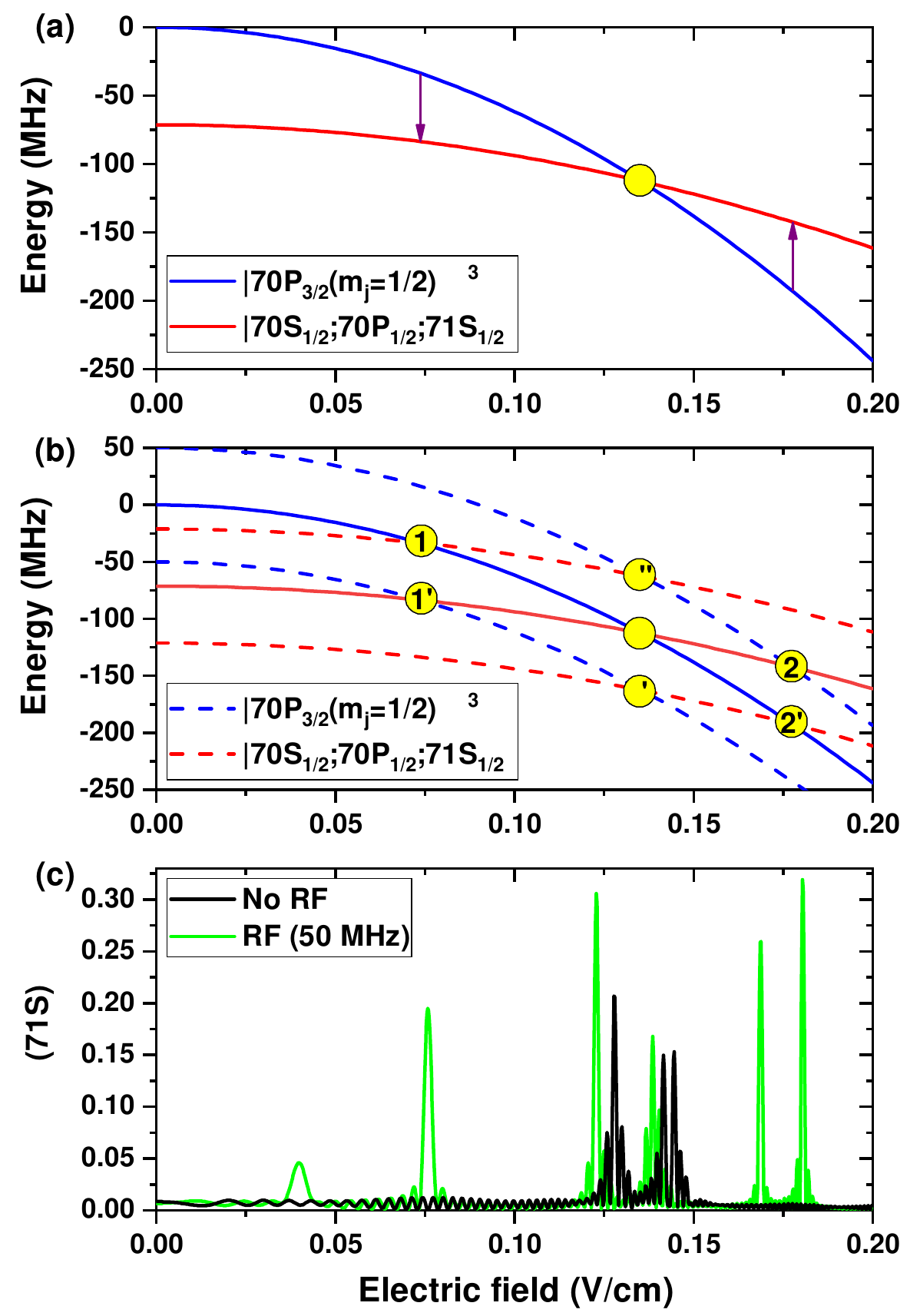}
\vspace{-.5cm}
\caption{
\label{Res1}
(a) Numerically calculated Stark structure of the collective energy levels, involved in three-body F\"{o}rster resonance (Eq. \ref{eq5}). Solid lines correspond to the initial (blue) and final (red) states of the system. The zero-field energy of the initial state $\ket{70P_{3/2}}^{\otimes 3}$ is taken as a reference. The intersection (I) marks the position of the three-body resonance (Eq. \ref{eq5}), while the arrows indicate the 1$^{st}$-order RF-induced resonances.
(b) Floquet representation of the Stark structure of collective energy levels. Dashed lines denote the 1$^{st}$-order Floquet sidebands of the initial (blue) and final (red) states. The intersections (I, I') and (I'') mark the position of the original three-body resonance (Eq. \ref{eq5}), while the intersections (1, 1') and (2, 2') indicate the corresponding 1$^{st}$-order RF-induced resonances.
(c) Numerically calculated dependence of the fraction $\rho$ of atoms in the final $\ket{71S_{1/2}}$ state caused by three-body resonance (Eq. \ref{eq5}). Central doublet represents original three-body resonance, while RF-induced resonances are represented by left- and right-hand side doublets. System parameters: $R=10$ $\mu$m, $F_{RF} = 0.05$ V/cm, $\nu = 50$ MHz, $T_{int} = 0.635$ \textmu s. The resonance marked with asterisk ($F_{S} = 0.1805$ V/cm) is used in further gate simulations.  
}
\end{figure}
\end{center}

Figure~\ref{Res1}(a) depicts the three-body Förster resonant transitions in a form of Stark diagram, calculated for non-interacting atomic system. The original resonance (Eq. \ref{eq5}) is represented by the intersection of solid blue (initial $\ket{70P_{3/2}}^{\otimes 3}$ state) and red (final $\ket{70S_{1/2};70P_{1/2};71S_{1/2}}$ state) lines, and is indicated by the symbol (I). This resonance naturally exists in the corresponding DC electric field ($F_S=0.135$ V/cm) even in absence of RF radiation. In turn, upon activation of the AC electric field component, one can observe the occurrence of additional 1$^{st}$-order induced resonance transfers displayed by the violet arrows in Fig.~\ref{Res1}(a). Thus, the left ($F_S = 0.074$ V/cm) and right ($F_S = 0.178$ V/cm) arrows correspond to the emission and absorption of an RF photon with frequency $\nu=50$ MHz, respectively. Note that resonances of higher orders can also arise with absorption/emission of a larger number of photons. Nevertheless, such resonances possess essentially smaller amplitudes (see Eq. \ref{eq4}) that, together with limited strength of three-body interactions, makes them poorly suitable for quantum operations realization \cite{Tretyakov2014, Yakshina2016}. We keep the modulation depth on the order of 1 during the simulations in order to get a strong $1^{st}$-order sideband and avoid unwanted spurious effects due to further sidebands.

\label{Sec3}
\vspace{-.5 cm}

Alternatively, we can describe the RF induction process using the Floquet approach. Thus, Figure~\ref{Res1}(b)  shows the initial and final collective Rydberg states, accompanied by the corresponding Floquet sidebands. Intersections (1) and (1') correspond to the left arrow in Fig~\ref{Res1}(a), while intersections (2) and (2') correspond to the right arrow. In turn, intersections (I') and (I'') emerge for the same DC field value as the original resonance (I). Notably, when considered in dressed state formalism, the radio-frequency sidebands only differ by the number of RF photons included. Thus, when found at the same DC field, the sidebands crossings represent the same resonant transfers and do not create additional interaction channels.

Figure~\ref{Res1}(c) presents the numerically calculated dependence of the fraction $\rho$ of atoms found in $\ket{71S_{1/2}}$ state on the DC electric field, thus demonstrating the efficiency of the three-body Förster resonant transfer (Eq. \ref{eq5}). When RF radiation is turned off, only original resonance (I) remains (black line). Note that in the resonant process, we cannot attribute various two-step transfers $\ket{70P_{3/2}} \to \ket{70S_{1/2}}$, $\ket{70P_{3/2}} \to \ket{71S_{1/2}}$ and $\ket{70P_{3/2}} \to \ket{70P_{1/2}}$ to a specific atom in the ensemble. In this regard, several resonant interaction channels are formed, from which only two are allowed due to symmetry reasons \cite{Ryabtsev2018}. Ultimately, this process is similar to the Autler–Townes effect, and leads to the splitting of resonant peaks into 2 satellites, as it is shown in Figure \ref{Res1}(c). Alternatively, when the RF radiation is present in the system, it induces additional first-order peaks (green line). The relative distances between the resonances correspond to the applied radiation frequency of 50 MHz. The displacement of the doublet centers relative to the expected resonant positions provided in Figures~\ref{Res1}(a,b), is due to the presence of the interatomic DDI taken into account in the complete simulation. The expected DC field of the resonance peak is also dependent on the AC Stark shift produced by the RF field applied. This effect is clearly noticeable for the original three-body resonance peak (black line in Fig.\ref{Res1}(c)), which shifts when the AC field component is switched on (central doublet, green line in Fig.\ref{Res1}(c)).

The slight difference in the shape and amplitude of the central peak in the presence/absence of RF radiation (Fig.\ref{Res1}(c)) is due to the difference in the interaction strength for these two cases. The population oscillation frequency of the DC-induced resonance ($\Omega_{DC}$) exceeds that of the RF-induced counterpart ($\Omega_{RF}$). In turn, the interaction time value $T_{int}=0.635$ \textmu s applied during simulation was chosen to maximize the amplitude of the RF-induced peaks under study, providing the pulse area $\Omega_{RF} T_{int} = \pi$. Accordingly, the pulse area of the DC-induced peaks $\Omega_{DC} T_{int} > \pi$, thus leading to the resonance amplitude reduction.

A great decrease of intensity can be seen for the left sideband (two leftmost peaks in Fig.\ref{Res1}(c), green line), when compared with the right sideband. This decrease is associated with two effects. First, the amplitude is influenced by two-body exchange interactions between atoms in $S$ and $P$ states, accompanied by changes of their momentum projections $m_j$. These interactions are resonant in the zero field due to the spectra degeneracy, thus provoking a significant leakage of the initial state population. The influence of such processes decreases quadratically when the DC field increases and becomes negligibly small for the original resonance \cite{Cheinet2020}. Second, the modulation depth (or equivalently the amplitude of the Floquet sideband) depends on the DC electric field. Since the Stark effect is quadratic, the modulation depth increases fast with the DC field component (see Eq. \ref{eq4}). Thus, a significant decrease in peak amplitudes can be expected for small values of $F_S$ compared to larger values. To avoid the unwanted influence of these two effects, we opt for the transitions arising for the DC field range $0.15 \leq F_{S} \leq 0.2$ V/cm when considering any resonant dynamics.

\begin{center}
\begin{figure}[!t]
\center
\includegraphics[width=\columnwidth]{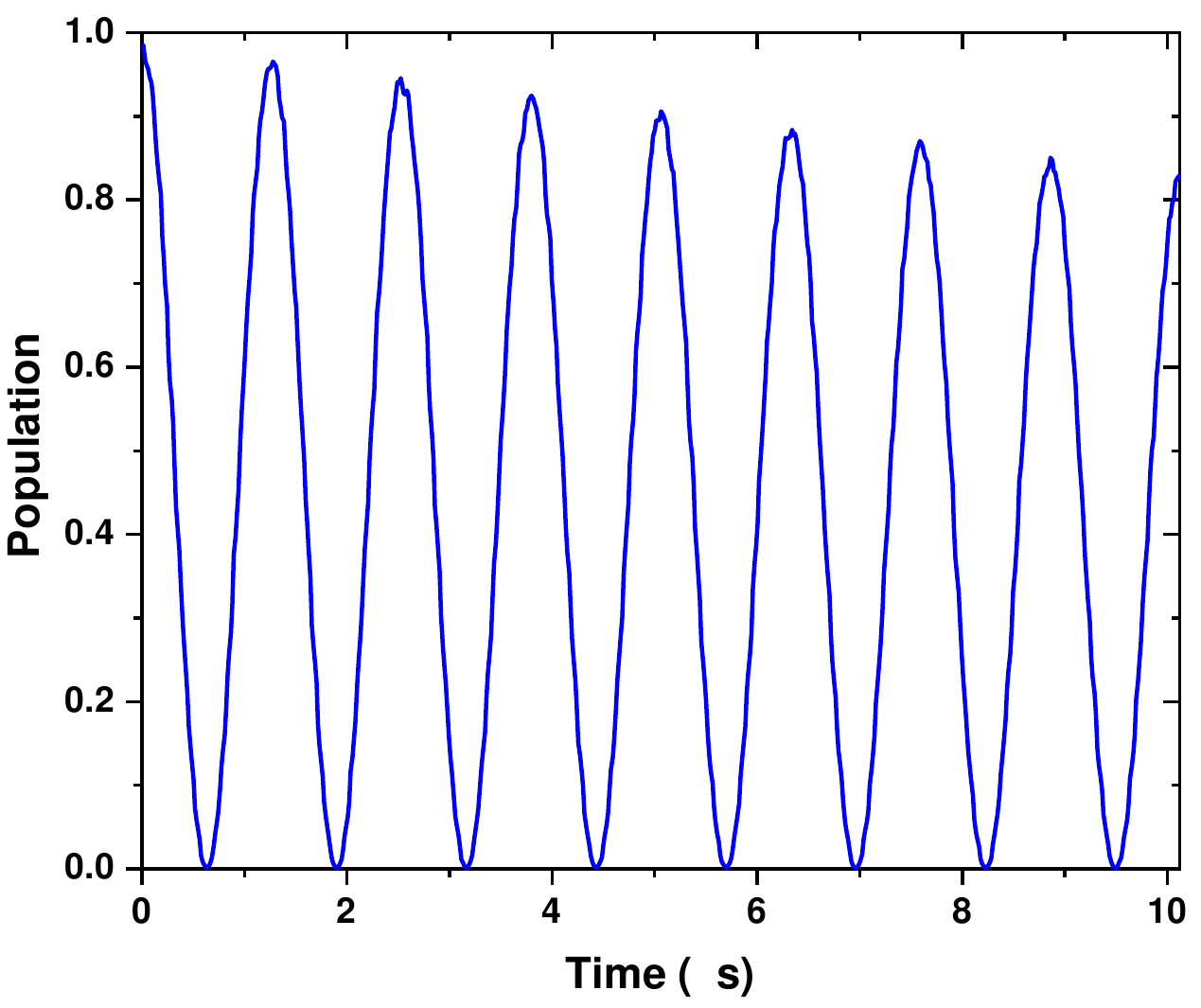}
\vspace{-.5cm}
\caption{
\label{Rabi}
Time dependence of the population $\Pi$ of the initial Rydberg state $\ket{70P_{3/2}}^{\otimes 3}$ during RF-induced three-body resonance (Eq. \ref{eq5}). System parameters: interatomic distance $R=10$ \textmu m; DC electric field value $F_S=0.1805$ V/cm; AC field amplitude $F_{RF} = 0.05$ V/cm; RF frequency $\nu = 50$ MHz, temperature $T=300$ K.
}
\end{figure}
\end{center}

When tuning the DC field to the rightmost resonant sideband $F_S = 0.1805$ V/cm (marked with asterisk in Fig.\ref{Res1}(c)), we observe coherent Rabi-like population dynamics of the collective Rydberg states (Fig.\ref{Rabi}). The oscillation period of $T_{osc}=1.27$ \textmu s corresponds to the RF-induced interaction strength. The amplitude decrease is mainly caused by the limited lifetimes of the Rydberg states. Note that the Rabi oscillations are also accompanied by fast phase dynamics, which can be directly controlled by the external field adjustment (see Sec. \ref{Sec4A}).

\vspace{-1.5 cm}

\vspace{.5 cm}
\section{General $CC\Phi$ gate protocol}
\label{Sec4}

RF-induced resonant interactions depicted in Fig.\ref{Res1} allow the direct control of phase and population dynamics of three-body Rydberg states in atomic ensemble. Thus, the presented resonances can be considered as promising candidates for the implementation of three-body quantum gate schemes. In this section, we present a quantum $CCPHASE$ gate (abbreviated here as $CC\Phi$ gate) protocol for arbitrary value of phase $\phi$, based on RF-induced Förster resonant transfers.

The $CC\Phi$ gate is a generalization of the well-known $CCZ$ gate for an arbitrary phase $\phi$ of a final multiqubit state. If both control qubits are in the state $\ket{1}$, a transformation is applied to the target qubit that changes the phase difference between its logic states by $\phi$. The extensive applicability of this gate to quantum algorithm implementations (e.g., for QAOA \cite{Hill2021}, QPE \cite{RevModPhys.86.153} and QFT \cite{Coppersmith2002, Nielsen2011}) makes it highly demanded in modern atomic quantum computing.

%The gate protocol we present is a substantial modification of our previous proposal \cite{Ashkarin2022}. The same linear arrangement of atomic qubits is kept, which was considered in the previous section. We utilize the central atom as a target qubit, and two outer atoms as control ones. The significant improvement lies in the application of non-resonant dc electric field during the gate process, which allows us to limit unwanted interactions during laser excitation, as well as to avoid the influence of always-resonant zero-field interactions. We use RF field to induce the F\"{o}rster resonance (Eq. \ref{eq5}) for a chosen value of $F_S$ and thus implement a quantum doubly-controlled rotation gate $CC\Phi$. Note that the dc part of the electric field is being kept constant during the whole process, thus facilitating the experimental implementation of the scheme.

The gate protocol we present is a substantial modification of our previous proposal \cite{Ashkarin2022}. The same linear arrangement of atomic qubits is kept, which was considered in the previous section. We utilize the central atom as a target qubit, and two outer atoms as control ones. The significant improvement lies in the application of non-resonant DC electric field during the gate process, which allows us to limit unwanted interactions in the system. Thus, the DC field is specifically chosen to isolate the RF-induced channel from unwanted non-resonant and quasi-resonant interactions in the atomic ensemble, which can negatively affect the laser excitation coherence \cite{Cheinet2020}. The influence of zero-field-resonant $S-P$ exchange interactions, which imply the change of $m_j$ is also limited. We use RF field to induce the F\"{o}rster resonance (Eq. \ref{eq5}) for a chosen value of $F_S$ and thus implement a quantum doubly-controlled rotation gate $CC\Phi$. Note that the DC part of the electric field is being kept constant during the whole process, thus facilitating the experimental implementation of the scheme.

\begin{center}
\begin{figure}[!t]
\center
\includegraphics[width=\columnwidth]{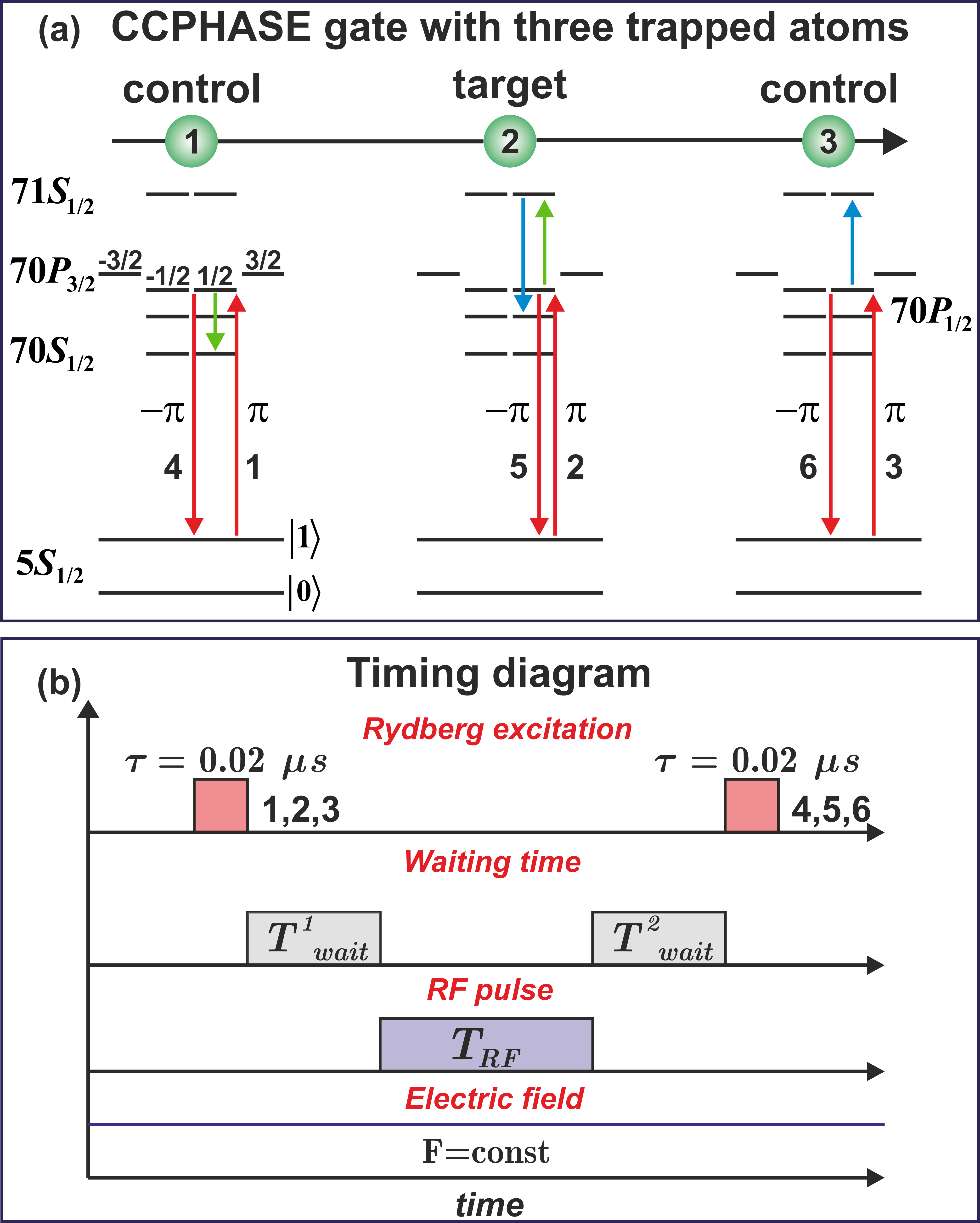}
\vspace{-.5cm}
\caption{
\label{Scheme}
(a) General scheme of the $CC\Phi(\phi)$ gate protocol based on three-body F\"orster resonance (Eq. \ref{eq5}). Three $^{87}$Rb atoms are located in individual optical tweezers aligned along the Z axis, which is co-directed with the external control electric field. Laser pulses 1-6 excite and deexcite the chosen Rydberg atomic states. The $\phi$ phase shift due to the three-body interaction appears only if all three atoms are excited into Rydberg states. The green and blue arrows here indicate $\ket{70P_{3/2}}^{\otimes 3} \to \ket{70S_{1/2};71S_{1/2};70P_{3/2}}$ and $\ket{70S_{1/2};71S_{1/2};70P_{3/2}} \to \ket{70S_{1/2};70P_{1/2};71S_{1/2}}$ two-body transitions, respectively. Thus, the full scheme corresponds to a three-body resonant transition (Eq. \ref{eq5}). Note that only one of several possible resonant energy transfers is shown. (b) Timing diagram of the pulses in the proposed gate scheme. Waiting times $T^{1(2)}_{wait}$ may be applied to fine-tune the relative phase of the various collective states.
}

\end{figure}
\end{center}

%=========================================================================================================
%=========================================================================================================
%==========================================================
\vspace{-.5 cm}

The implementation of the desired $CC\Phi$ gate can be divided into a sequence of three steps (see Fig.\ref{Scheme}).

Step 1: The laser excitation $\pi$ pulses 1-3 are applied simultaneously to all three atomic qubits. We assume that prior to the excitation, the atoms were in their logical states (namely, $\ket{0}$ or $\ket{1}$). These states are represented by two pre-selected hyperfine sublevels of rubidium ground state $\ket{5S_{1/2}}$ \cite{Saffman2010, Beterov2021}. We consider fully resonant single-photon laser excitation in rotating wave approximation, assuming that only transitions of the form $\ket{1} \to \ket{R} = \ket{70P_{3/2}\left(m_j=1/2\right)}$ are allowed. For practical applications, a three-photon excitation scheme can be used to pair the logical states of qubits with Rydberg levels~\cite{Ryabtsev2011, Beterov2023}. The effects associated with the phase and intensity noise of the laser were considered in \cite{De2018} and are not taken into account in this study.

Step 2: An RF pulse of duration $T_{RF}$ is applied to the system, inducing resonance for a certain value of the DC electric field. This pulse is accompanied by two waiting times $T_{wait}^1$ and $T_{wait}^2$, which are as well configurable system parameters. The DC and AC parts of the electric field are co-directed, thus keeping the symmetry of the atomic system unchanged. The resonant interaction causes the phase of the collective state $\ket{RRR}$ to change by $\phi$ during the gate time. This corresponds to a $CC\Phi(\phi)$ gate implementation, provided that all other states of the system remain unchanged after the application of RF radiation is terminated.

Step 3: The de-excitation $-\pi$ pulses 4-6 are applied simultaneously to all three atomic qubits. The pulses phase values are inverted compared to excitation pulses 1-3 in order to prevent additional unwanted phase accumulation. The $\phi$ phase difference is thus mapped from collective Rydberg states into the logical states of the target qubit.

\section{CCZ gate}
\label{Sec4A}

\begin{center}
\begin{figure*}[t!]
\center
\includegraphics[width=\textwidth]{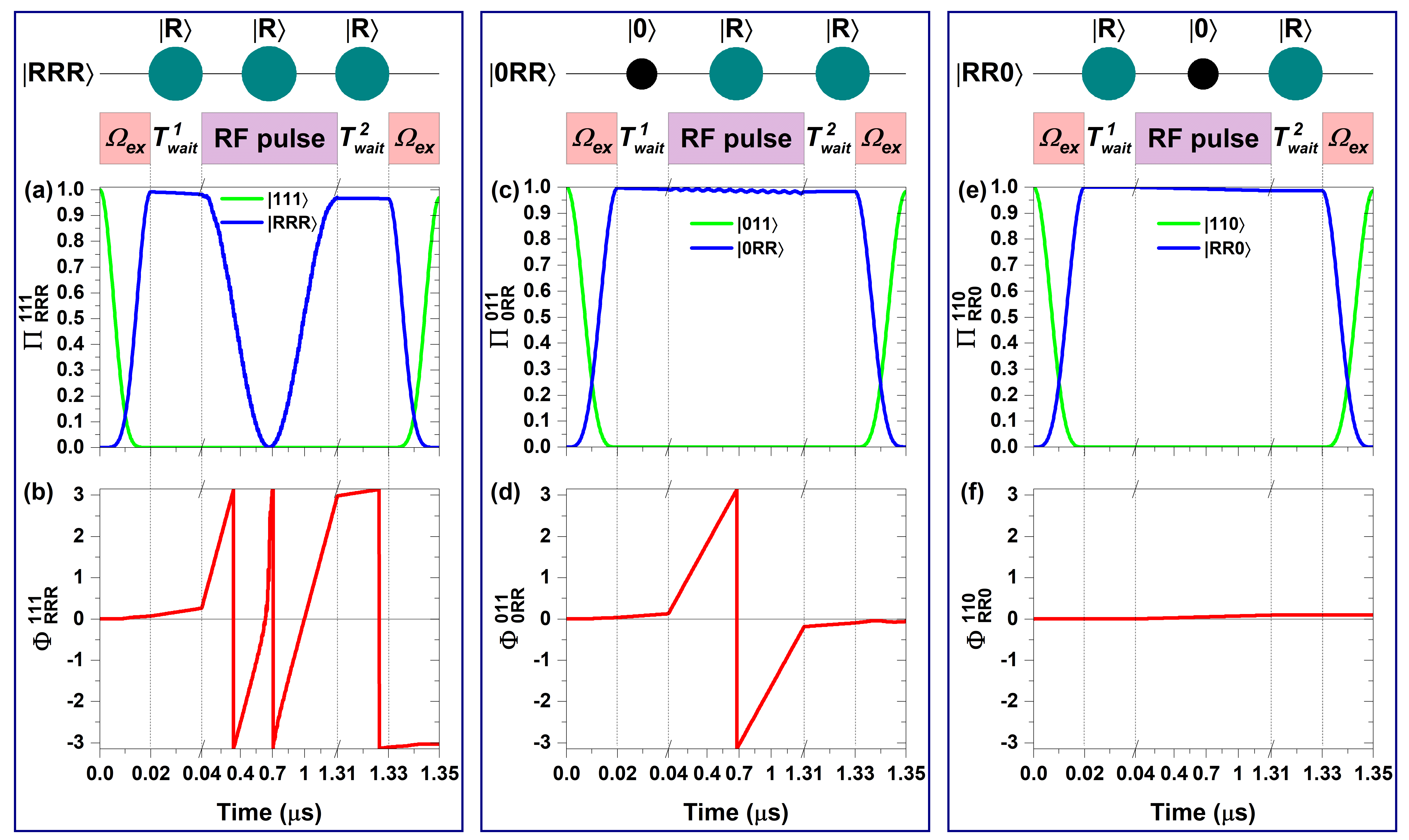}
\vspace{-.5cm}
\caption{
\label{CCZ}
Numerically calculated population and phase dynamics during $CC\Phi (\pi)$ gate implementation. (a) Time dependencies of the $\ket{111}$ initial logical state (green) and the corresponding $\ket{RRR}$ Rydberg state (blue) populations. Strong three-body resonance (Eq. \ref{eq5}) is visible during the RF pulse application. (b) Weighted phase evolution (modulo  $2\pi$) for the initial register state $\ket{111}$. Phase is changed by $\pi$ after the gate execution. (c) Population dynamics for the $\ket{011}$ ($\ket{101}$) initial logical state and Rydberg $\ket{0RR}$ ($\ket{R0R}$) state, corresponding to the case of excitation of two nearby atoms in the ensemble. (d) Weighted phase evolution for the initial register state $\ket{011}$ ($\ket{101}$). Phase change is compensated to zero during the gate process. (e) Time dependencies of the $\ket{110}$ initial logical state and $\ket{RR0}$ Rydberg state populations, corresponding to the case of excitation of two outermost atoms. (f) Weighted phase evolution for the initial register state $\ket{110}$. No sizable phase evolution is presented. System parameters: interatomic distance $R = 10$ \textmu m; DC electric field $F_S=0.1805$ V/cm; RF field amplitude $F_{RF}=0.05$ V/cm; RF frequency $\nu = 50$ MHz; RF pulse duration $T_{RF}=1.27$ \textmu s; waiting times $T^1_{wait}=T^2_{wait}=20$ ns; excitation/deexcitation times $T_{ex}=T_{deex}=20$ ns; temperature $T=300$ K. The estimated gate fidelity can be seen in Table \ref{tab}.
}
\end{figure*}
\end{center}
\vspace{-.7 cm}

We performed numerous simulations of the $CC\Phi(\phi)$ gate protocol for a wide range of $\phi$. The simulations were implemented in an extended basis, which also included the logical states of qubits in order to take into account the Rydberg interactions already arising during the excitation step. For the first demonstration of the protocol operation, we consider the implementation of the $CC\Phi(\pi)$ gate, commonly referred to as $CCZ$ gate. Results for several alternative values of $\phi$ are shown in Section \ref{AppA}.

The population and phase dynamics of the three-atom system during gate operation is shown in Figure \ref{CCZ}. To perform a gate according to the protocol described above, we use the asterisk-marked Floquet sideband shown in Figure \ref{Res1}(c). Thus, the atoms are excited into Rydberg states $\ket{70P_{3/2}}$ in an external DC field $F_S=0.1805$ V/cm. Then, the resonant interaction is activated by applying the inducing RF radiation of $\nu = 50$ MHz. In order to demonstrate the relevant phase dynamics over the whole gate sequence, we describe the evolution of the weighted phase $\Phi_{R_i}^{l_i}$ defined by Eq. (\ref{eq6}). Here $l_i$ is one of the register collective logical states, $R_i$ is the Rydberg state, which is coupled with $l_i$ by the excitation pulses 1-3, and $a_{i}$ denotes the complex amplitude of the corresponding state. Weighted phase allows one to accommodate the phase changes of both logical and Rydberg qubit states within a single meaningful smoothly changing variable, providing a convenient analytical tool.
\begin{equation}
    \label{eq6}
\Phi_{R_i}^{l_i}=\frac{|a_{R_i}|^2 \phi_{R_i}+|a_{l_i}|^2 \phi_{l_i}}{|a_{l_i}|^2+|a_{R_i}|^2}
\end{equation}
Note that here we use standard qubit state notation order for the register states, related to the qubit role in the gate and not to its physical location in the array. Thus, a $\ket{Control} \otimes \ket {Control} \otimes \ket{Target}$ state representation is used, although in the described gate configuration the target qubit is the central one.

If all three atoms were initialized in logical state $\ket{1}$, the collective Rydberg state $\ket{RRR}$ is excited during the time $T_{ex}$ (represented by left $\Omega_{ex}$ symbol in Fig.\ref{CCZ}(a)). In this case, we observe resonant Rabi-like population oscillations (Fig.\ref{CCZ}(a)), accompanied by fast phase dynamics (Fig.\ref{CCZ}(b)). Additional slow phase evolution also occurs during the excitation and deexcitation steps, as well as during the waiting times $T^{1(2)}_{wait}$. We optimize the waiting time periods such that the overall evolution results in a $\pi$ phase change after the interaction time. This corresponds to the controlled phase shift for the initial state $\ket{111}$ in Fig.\ref{Scheme}(a). This phase shift is sensitive to both DC electric field and RF pulse amplitude, which act directly on the F\"orster defect. The population of initial state after the RF-induced resonance is $96.2\%$ due to the finite Rydberg lifetimes and the leakage of population to other collective levels by Rydberg interactions. These effects are found to be the major sources of the gate error.

Alternatively, if two atoms out of three are excited into Rydberg states, only off-resonant two-body interactions $\ket{70P_{3/2}\left(m_j=1/2\right)}^{\otimes 2}\to \ket{70S_{1/2}\left(m_j=1/2\right);71S_{1/2}\left(m_j=1/2\right)}$ are possible due to the selection rule $\Delta M=0$. As discussed in \cite{Ashkarin2022}, these interactions only lead to substantial dynamics for closely-spaced atoms due to the fast decrease of van der Waals interaction strength with distance. Thus, for state $\ket{RR0}$ (coupled to the register state $\ket{110}$), which correspond to the case of excitation of two corner atoms, the population dynamics is mainly limited by the Ryderg state decay, accompanied by modest phase evolution (see Figs.\ref{CCZ}(e,f))  Nevertheless, for states $\ket{0RR}$ and $\ket{R0R}$ one can observe fast phase evolution provoked by off-resonant two-body transitions (Fig.\ref{CCZ}(d)). Note that since these two states are analogous in chosen register configuration, they experience the same time dynamics. The accumulated phase shift is sensitive to the external electric field and can be compensated to zero during the interaction time $T_{RF}+T^1_{wait}+T^2_{wait}$, taking into account the slow phase dynamics for partially-excited Rydberg states and additional evolution during the waiting times. Two-body interactions also have a residual impact on the population of the initial state, leading to weak Rabi-like oscillations (amplitude $\sim1-2\%$) (Fig.\ref{CCZ}(c)). A detailed analysis of two-body interactions in three-body systems of Rydberg atoms is given in our previous papers~\cite{Ashkarin2022, Cheinet2020, Ryabtsev2018, Beterov2018}.

Finally, when only one atom in the ensemble is temporarily excited (forming Rydberg states $\ket{00R}$, $\ket{0R0}$ or $\ket{00R}$), the $\pi$ and $-\pi$ pulses, shown in Fig.\ref{Scheme}(a), will return the system into the initial state with zero phase shift after the interaction time $T_{RF}+T^{1}_{wait}+T^{2}_{wait}$. However, temporary Rydberg excitation will result in population loss due to the finite lifetimes of Rydberg states, with the loss magnitude dependent on the environmental temperature $T$. The trivial case is when no Rydberg atoms are excited (state $\ket{000}$). We assume that the pulses 1-6 will have no effect in this instance, leaving the ground states unperturbed. 

Considering the phase and population dynamics described above, we come to the following conclusions: as a result of the algorithm implementation, the phase of the $\ket{111}$ state changes by $\pi$ during the gate execution time (see Fig.\ref{CCZ}(b)). In contrast, the phase shifts for the other basis states of the quantum register remain negligibly small, or are compensated for by weak non-resonant interactions. Thus, the algorithm performs the quantum gate operation $CC\Phi(\pi) \equiv CCZ$.

To perform additional verification of the protocol functionality, we numerically simulated the Toffoli gate scheme implementation based on proposed $CCZ$ gate. We apply two additional Hadamard gates to the target qubit, thus constructing the doubly-controlled-NOT operator. These Hadamard gates can be performed by two-photon Raman $Y_{\pi/2}$ pulses \cite{Zeng2017}, and are considered perfect during the gate simulation.

\begin{center}
\begin{figure}[!ht]
\center
\includegraphics[width=\columnwidth]{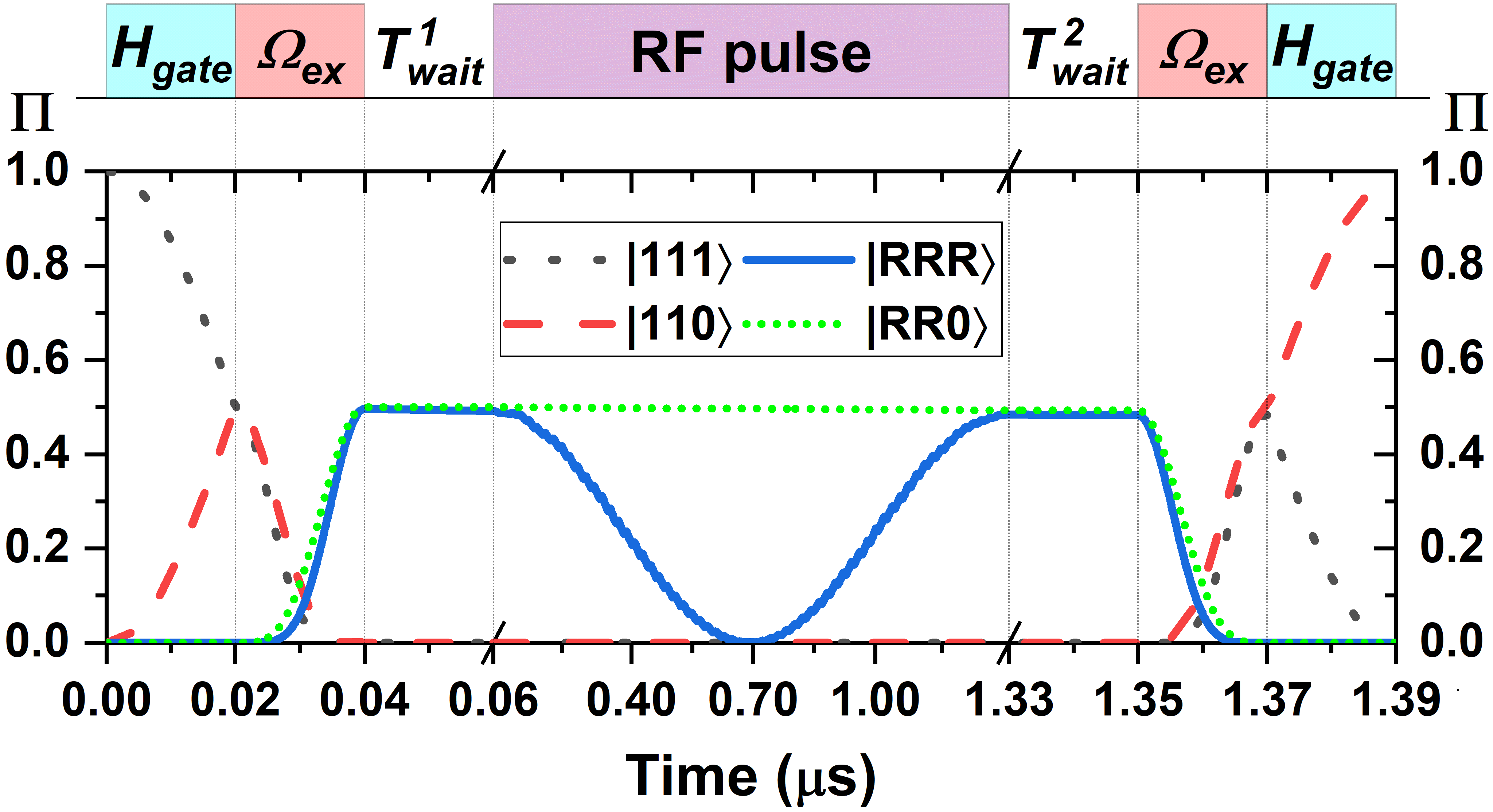}
\vspace{-.5cm}
\caption{
\label{Toffoli}
Population dynamics of collective states of a triatomic register during the Toffoli gate implementation based on the RF-induced three-body resonance (Eq. \ref{eq5}). The applied pulses are indicated at the top of the figure. The initial state of the quantum register is represented by a black dotted line, and the final state is represented by a red dashed line. The blue and green lines denote the corresponding collective Rydberg states. System parameters: $R = 10$ \textmu m; $F_S=0.1805$ V/cm; $F_{RF}=0.05$ V/cm; $\nu = 50$ MHz; $T_{RF}=1.27$ \textmu s;  $T^1_{wait}=T^2_{wait}=20$ ns; $T_{ex}=T_{deex}=20$ ns; $T=300$ K.
}
\end{figure}
\end{center}

Figure~\ref{Toffoli} shows the behavior of the relevant collective states of the system in the case when the initial state of the register is $\ket{111}$. Due to the application of the Hadamard gate, the population is first evenly distributed between states $\ket{111}$ and $\ket{110}$. Then the atoms are excited into the Rydberg state $\left(\ket{RRR} + \ket{RR0}\right)/ \sqrt{2}$. The subsequent RF pulse results in the application of $CCZ$ gate, shown in Fig.\ref{CCZ}, thus leading to the final population interchange between $\ket{111}$ and $\ket{110}$ states. The stages of de-excitation of atoms and the second application of the Hadamard gate complete the gate protocol.

\vspace{-1 cm}

\section{Arbitrary $CC\Phi(\phi)$ gates}
\label{AppA}

\begin{center}
\begin{figure*}[!hbt]
\center
\includegraphics[width=\textwidth]{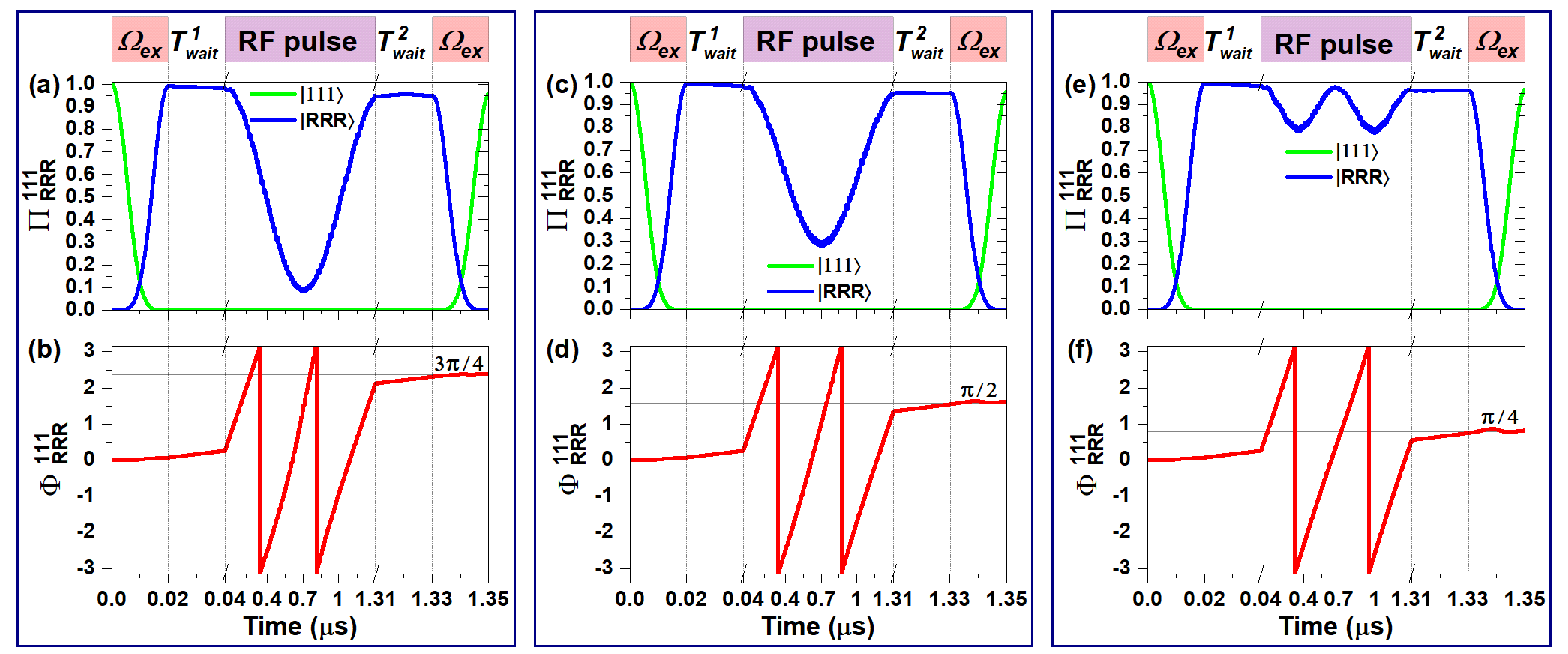}
\vspace{-.5cm}
\caption{
\label{CCPh}
Numerically calculated dynamics of populations and weighted phases of logical register state $\ket{111}$ and corresponding Rydberg state $\ket{RRR}$ during the implementation of $CC\Phi(\phi)$ gates for different $\phi$. (a,b) Population and phase dynamics during the gate implementation for $\phi = 3\pi/4$. Slightly off-resonant population behaviour is observed, accompanied by a slowed phase evolution, when compared with the $CCZ$ gate case (see Fig.\ref{CCZ}(a,b)). RF radiation parameters: $F_{RF}=0.0416$ V/cm; $\nu = 48.35$ MHz. (c,d) Analogous population and phase dynamics for the $CC\Phi(\pi/2)$ gate implementation. Increased resonance detuning results in the reduction for phase accumulation rate. RF radiation parameters: $F_{RF}=0.03409$ V/cm; $\nu = 47.1$ MHz. (e,f) Population and phase evolution for the $CC\Phi(\pi/4)$ gate implementation. Strongly off-resonant transfer leads to deceleration of phase accumulation, when compared with the previous cases. RF radiation parameters: $F_{RF}=0.0389$ V/cm; $\nu = 46.75$ MHz. Note that the simulation results for $T=300$ K are presented. The estimated gate fidelities can be seen in Table \ref{tab}.
}
\end{figure*}
\end{center}

\vspace{-.5 cm}

Section \ref{Sec4A} described the implementation of $CC\Phi (\pi) \equiv CCZ$ gate in a linear $^{87}$Rb quantum register. In this section, we present exemplar numerical implementations of the $CC\Phi(\phi)$ gate for alternative values of the phase $\phi = \{\pi/4, \pi/2, 3\pi/4\}$. During the simulations, we aimed to achieve the maximum gate fidelity while keeping all basic experimental parameters unchanged compared to $CC\Phi (\pi)$ gate protocol. We leave fixed the interatomic distances $R=10$ \textmu m in the register, thus effectively retaining the interaction strength. Accordingly, the gate for any phase value is realised using the asterisk-marked Floquet peak in Figure \ref{Res1}(c), at a fixed value of the DC electric field $F_S=0.1805$ V/cm. To adjust the population and phase evolution patterns, we vary the frequency of inducing RF radiation $\nu$ and its amplitude $F_{RF}$. Note that all the timings are also preserved during the gate implementation.

Figure \ref{CCPh} illustrates the phase and population dynamics of the quantum register states during the $CC\Phi(\phi)$ gate implementation. Since the evolution of the system in the case of two-atom excitation does not change noticeably for different values of the accomodated gate phase, the dynamics of the $\ket{0RR}$ ($\ket{R0R}$) and $\ket{RR0}$ states depicted in Figs.\ref{CCZ}(c,d,e,f) remain relevant for all the gates presented in this section. Thus, Fig.\ref{CCPh} only depicts the evolution of the logical state $\ket{111}$ and its Rydberg coupled state $\ket{RRR}$.

In Figure \ref{CCPh}(a), the time dependencies of the logical and Rydberg states populations during the $CC\Phi(3\pi/4)$ gate implementation are shown. We observe a significant decrease in the amplitude of Rabi oscillations during the RF-induced transition for the $\ket{RRR}$ state due to the presence of a weak detuning from the exact resonance necessary to slow down the phase dynamics. As a result of this slowdown, phase evolution allows to reach a value of $\phi=3\pi/4$ during the protocol implementation period (see Fig.\ref{CCPh}(b)). 

Increasing the resonance detuning by varying the RF radiation parameters, we slow down the phase evolution rate during the three-body transfer. Thus, for the population dynamics during the $CC\Phi(\pi/2)$ and $CC\Phi(\pi/4)$ gate implementations shown in Figs.\ref{CCPh}(c,e), the Rabi oscillations amplitude decrease is clearly noticeable, which lead to a significant phase rate decrease (Figs.\ref{CCPh}(d,f)).

Note that the described gate protocol can be realised for arbitrary atomic system parameters. Thus, if certain values of the principal quantum number $n$, interatomic distance $R$ and DC electric field $F_S$ have been chosen due to experimental requirements, one will be able to select a suitable radio-frequency pulse to realize the $CC\Phi(\phi)$ gate, given that strong enough three-body transfer can be induced by RF radiation.

\vspace{.5 cm}
\section{Gate fidelity and optimization}
\label{Sec4B}

To estimate the individual gate fidelity, the method proposed in~\cite{Bowdrey2002} is used. We consider 6 single-qubit configuration states: $\ket{0}$, $\ket{1}$, $\left( \ket{0}+\ket{1} \right)/ \sqrt{2}$, $\left( \ket{0}-\ket{1} \right)/ \sqrt{2}$, $\left( \ket{0}+i\ket{1} \right)/ \sqrt{2}$ and $\left( \ket{0}-i\ket{1} \right)/ \sqrt{2}$. We form a set of three-qubit states as all $6^3=216$ combinations of three single-qubit basis states. We simulate the density matrices $\rho_{sim}$ of all final states after $CC\Phi(\phi)$ gate is applied to each initial state. Then we calculate the fidelity of each final state comparing to the reference state $\rho_{et}$, which is the final state of the ensemble after the perfect $CC\Phi(\phi)$ gate is performed~\cite{Nielsen2011}:
\begin{eqnarray}
\label{eq7}
F=\textrm{Tr}\sqrt{\sqrt{\rho_{et}}\rho_{sim}\sqrt{\rho_{et}}}
\end{eqnarray}
Averaging over all 216 states, we calculate $CC\Phi(\phi)$ gate fidelities for different values of $\phi$. Several exemplar results are shown in Table \ref{tab}. Note that the characteristics of the applied RF radiation are the only controlling parameters of the gate. To perform different phase gates, we only change $F_{RF}$ and $\nu$, and keep all the other parameters fixed (see Sec. \ref{AppA}).

\begin{table}
\caption{\label{tab} Estimated $CC\Phi(\phi)$ gate fidelities for different values of $\phi$. Results for room temperature ($T=300$ K) and cryogenic setup ($T=4$ K) are presented. System parameters: $R=10$ \textmu m, $F_S=0.1805$ V/cm, $T_{RF}=1.27$ \textmu s, $T_{wait}^1=T_{wait}^2=T_{ex}=T_{deex}=20$ ns. RF parameters are shown in the table for each gate.}
\begin{ruledtabular}

\begin{tabular}{c c c c c}
 $T$ (K)&$CC\Phi(\frac{\pi}{4})$ & $CC\Phi(\frac{\pi}{2})$ & $CC\Phi(\frac{3\pi}{4})$ & $CC\Phi(\pi)$ \\
 300 & 99.27\% & 99.26\% & 99.22\% & 99.31\% \\
 4 & 99.64\% & 99.63\% & 99.6\% & 99.69\% \\
 \hline
 $F_{RF}$ (V/cm) & 0.0389 & 0.03409 & 0.0416 & 0.05 \\
 $\nu$ (MHz) & 46.75 & 47.1 & 48.35 & 50 \\
\end{tabular}
\end{ruledtabular}
\end{table}

For the described experimental conditions of resonance (Eq. \ref{eq5}), the average fidelity of the gate implementation is $99.27\%$ for the room-temperature environment. The main source of fidelity losses ($\sim 0.51\%$) is the finiteness of the lifetimes of Rydberg states. Compensation for these losses down to $0.13\%$ can be achieved using a cryostat at 4 K \cite{Ximenez2023, Schymik2021} temperature for the proposed parameter values, increasing the average fidelity to $99.65\%$. Additional compensation for Rydberg lifetime losses can be achieved by using higher Rydberg levels \cite{Wu2021a, Saffman2010}. The losses associated with the non-optimal choice of parameter values are estimated as $\sim 0.22\%$ after the application of multiparametric optimization routines based on simulated annealing technique \cite{Suman2006}. However, this optimization is preliminary and does not guarantee to have reached the global optimum for parameter values due to the complexity of the model. Remaining losses could be compensated for by applying additional multi-parametric optimization routines. For practical applications, we propose to use simulated annealing method in conjunction with Nelder-Mead pre-processing \cite{Ali2014} or GRAPE-based optimization routines \cite{Jandura2023}. QAOA optimization techniques are also applicable \cite{Dlaska2022, Nguyen2023}.

To implement correct phase gates experimentally, it is necessary to pay attention to the required accuracy of the parameter values control. Assuming that the allowable fidelity deviation from the maximum value cannot exceed $0.1\%$ we found the following requirements for parameter accuracy thresholds: the interatomic distance must be controlled with an accuracy of $20$ nm; the interaction time~-~$0.025$ \textmu s; DC electric field amplitude~-~$7 \cdot 10^{-5}$ V/cm, AC electric field amplitude~-~$9.2 \cdot 10^{-5}$ V/cm, RF frequency - $20$ kHz. While the control of the interaction time, as well as of the RF radiation frequency, does not present any experimental challenges \cite{Vankka2001}, the limiting factors are presented by the complexity of the DC and AC field amplitudes control, along with the interatomic distance control. As shown in our previous work \cite{Ashkarin2022}, one can use closer interatomic distances for the chosen value of $n$ to reduce the possible errors in field amplitude control. Nevertheless, this proposal implies an interchange between different error sources, depending on the experimental requirements, and could be more applicable for protocol implementation in optical-lattice-based quantum registers \cite{Beals2008}. A complete analysis on electric field recording was proposed both for DC \cite{Facon2016, Bowden2017} and AC \cite{Meyer2021, Sedlacek2012, Liao2020} fields in Rydberg systems. A proper level of interatomic distance control can be attained with modern holographic and AOD-based tweezer techniques \cite{Chew2022, Xu2022, Chen2022}. Appropriate composite pulse sequences \cite{Torosov2011, Torosov2011a} and coherent control techniques \cite{Warren1993, Larrouy2020} may also be applied to reduce sensitivity to the experimental parameters.

\section{Conclusion}
In this paper, we have proposed and numerically investigated a new kind of three-body Förster resonant interactions induced by radio-frequency radiation in Rb Rydberg atoms. We then proposed a novel three-qubit quantum phase gate protocol based on the investigated resonances. The high protocol fidelity ($\sim 99.27 \%$) is mainly limited by the finite Rydberg lifetimes and can be significantly increased (up to $\sim 99.65 \%$) in a cryogenic environment. The gate behaviour is controlled via the RF radiation parameters solely, which greatly facilitates the implementation of the protocol and makes it more robust to deviations of DC field amplitude when compared to our previous proposals \cite{Beterov2018a,Ashkarin2022}.

Based on the presented research, we conclude that RF-induced F\"{o}rster resonances provide ample opportunities for the implementation of quantum operations in registers of ultracold neutral atoms. RF induction allows versatile tuning of the positions and strengths of resonance peaks, opening access to high-precision Rydberg dynamics control and entanglement production in large-scale registers. The high flexibility of the resonant protocol allows it to be implemented in various quantum register architectures, paving the way for the creation of fully interconnected neutral-atom-based quantum devices.

\section*{Acknowledgement}
We thank Daniel Comparat for providing feedback on the manuscript. This work was supported by French government, under the  \enquote{Vernadski} scholarship program grant, and by French National Research agency (ANR), under grant ANR-22-CE47-0005 (QIPRYA project). The Russian team was supported by Russian Science Foundation grant No. 23-12-00067.

\newpage
\bibliography{JCbib_old1}{}

\end{document}